\documentstyle[12pt]{article}
\def\be{\begin{equation}}
\def\ee{\end{equation}}
\def\bea{\begin{eqnarray}}
\def\eea{\end{eqnarray}}
\def\a{\alpha}
\def\b{\beta}
\def\e{\varepsilon}
\def\g{\gamma}

\def\d{\delta}

\def\t{\theta}
\def\h{\eta}
\def\z{\zeta}
\begin{document}
\renewcommand{\theequation}{\thesection.\arabic{equation}}
\title{Gauss-Bonnet lagrangian  $G$ ln $G$ and cosmological exact solutions}
\author{Hans-J\"urgen  Schmidt}
\date{March 18, 2011}
\maketitle
\centerline{Institut f\"ur Mathematik, Universit\"at Potsdam, Germany} 
\centerline{Am Neuen Palais 10, D-14469 Potsdam,  \  hjschmi@rz.uni-potsdam.de}
\begin{abstract}
For the lagrangian $L =  G \ln  G$ where $G$ is the Gauss-Bonnet
curvature scalar we deduce the field equation and solve it in closed
form for 3-flat Friedman models using a statefinder parametrization. \par
Further we show, that among all lagrangians $F(G)$ this $L$ is the 
only one not having the form $G^r$ with a real constant $r$ but
possessing a scale-invariant field equation. This turns out to be one of its 
analogies to $f(R)$-theories in 2-dimensional space-time.  \par
In the appendix, we systematically list several formulas for the 
decomposition of the Riemann tensor in arbitrary dimensions $n$, which
 are  applied  in the main deduction for $n=4$.
\end{abstract}

\section{Introduction}%1
\setcounter{equation}{0} 
Fourth-order gravity has been a serious alternative to General Relativity since 1918  
already: H. Weyl, see \cite{m10}, was guided by the idea of the scale  invariance of the 
action which requested for an $R^2$-term in its integrand instead of the Einstein-Hilbert 
action integrand $R$. 
In fact, the integral $\int R^n \sqrt{-g} d^kx$ in $k$-dimensional space-time is 
scale-invariant just in the case $k = 2n$, leading to $n=2$ for the usual space-time 
dimension $k=4$. For   details see e.g. the reviews  \cite{sc136},   \cite{c80},
 \cite{d36}, and \cite{d37},   and the books  \cite{d38} and   \cite{d31}. 
For  a broader view to this topic, and also on the growth of  (quantum) perturbations 
to the today's observed large-scale structures by inflation,  see  the references cited  
there. \par
Since 1947 it became clearer, that the cosmological evolution can 
be better modeled if both $R$ and $R^2$-terms belong to  the action, see C.~Gregory 
 \cite{m11}. In the eighties, the inflationary cosmology has been related to 
fourth-order gravity  by   Starobinsky \cite{m12}, and this paper  initiated several 
follow-up papers, e.g.  \cite{sc19} and \cite{sc25}; generalizations by inclusion 
of  $R^3$-terms and  later by a general $f(R)$ have been worked out e.g. in  
 \cite{sc24} and \cite{sc32}.  \par
 In 1921,  R. Bach \cite{m13}, see  \cite{sc109} for details,  initiated a detailed
investigation of  the conformally invariant  field equations following from the lagrangian
$C_{ijkl} C^{ijkl}$. In 1977 it was shown, that a theory with lagrangian of  the form
\be%1.1
\Lambda + R + \a R^2 + \b C_{ijkl} C^{ijkl}
\ee
where units are chosen that light velocity equals $1$ and 
 Newton's  constant equals $(16 \pi)^{-1} $, can be
renormalized, see K. Stelle \cite{m14}. \par
In this context it is often mentioned, that the addition of  a multiple of the 
Gauss-Bonnet term, see also \cite{sc03} and  \cite{m15}, 
\be%1.2
G = R_{ijkl} R^{ijkl} - 4 R_{ij} R^{ij} + R^2
\ee
to a lagrangian like (1.1) does not alter the field equations, but it leads to a surface term 
which may become essential in the quantization.  The relation to topology is as
follows: the field equations come out by applying  continuous deformations of the
metric, but $\int G  \sqrt{-g} d^4 x$ is a topological invariant, see \cite{m16},
 \cite{m17},  \cite{c77}, and \cite{d34}. Lanczos deduced only the 4-dimensional case, 
whereas Lovelock 
generalized to arbitrary dimensions. His sequence $L_n$ starts with
$L_1=R$, $L_2=G$, and each $L_n$   leads to a topological invariant in 
the $2n$-dimensional space or space-time. \par
Recently, a lot of papers appeared which contain the Gauss-Bonnet term in the 
action. To circumvent the vanishing of its variational derivative, essentially
 three ways have been gone: Models in dimension larger than 4, see  e.g. \cite{c50},
\cite{b80},  \cite{d15}, \cite{my1}, \cite{my2}, 
 and  \cite{d32}, models where $G$ is multiplied by a 
scalar $\phi$, see  \cite{d19}, \cite{d35}, and models where $F(G)$ instead of $G$
 is used in the lagrangian with a suitably chosen  non-linear function $F$, see
e.g. \cite{d33}. Applications of  theories with Gauss-Bonnet term to cosmology
 can also be found in \cite{m18},  \cite{m19},  \cite{m20},  \cite{m21},  
\cite{m22},   \cite{mx18},  \cite{mx19},  \cite{mx20},  \cite{mx21}, 
\cite{mx22}, \cite{mx25}, \cite{mx26},  and  \cite{mx27}.   

\section{Statefinder parametrization}%2
\setcounter{equation}{0} 
The metric of a 3-flat Friedman model with synchronized time  coordinate $t$ reads
\be%2.1
ds^2 = dt^2 - a^2(t) \left(dx^2 + dy^2 + dz^2 \right)\, ,   \qquad a(t) > 0\,  .
\ee
We assume that the Taylor development of  $a(t)$ exists, the dot denotes 
$d/dt$, and the Hubble parameter $h$ is defined as usual  via
\be%2.2
h(t) = \frac{\dot a}{a} = \dot \a \, , \qquad \a = \ln a\, .
\ee

In what follows, we always exclude a constant function  $a(t)$ as it 
represents the trivial Minkowski space-time solution.  So, we restrict
to functions $a(t)$ which have $h(t) = 0$ at isolated moments of time 
only; at those moments, our exact solutions to be deduced below,
have to be matched together.  \par
A time-inversion leads to a change of the sign of $h$, so we may assume in 
the following, that always $h(t) > 0$. Under these circumstances we define
for any natural number $n \ge 2$
\be%2.3
z_n = \frac{a^{(n)} \, a^{n-1}}{\dot a^n}\, , \qquad a^{(n)} = \frac{d^n a}{dt^n}\, .
\ee
This expression is, up to a constant factor, uniquely determined by the 
conditions that it is proportional  to the $n$-th time derivative of $a$
with proportionality factor containing $a$ and $\dot a$ only,  
is time-reparametrization invariant, and  is scale-invariant. By obvious reasons 
it proves useful to define also $z_0 = z_1 =1$. \par
Another, but equivalent method to define the parameters $z_n$ goes as follows:
it is  the only  product  of  $a^{(n)}$  with powers of $a$ and $\dot a$  which is  
dimensionless in both interpretations of  metric (2.1). In the first interpretation of (2.1), 
$a$ is a  dimensionless quantity, dimensions   are encoded in $t$, $x$, $y$ and $z$. 
In the second interpretation of (2.1), $x$, $y$ and $z$   are dimensionless quantities, 
dimensions  are encoded in $t$ and $a$; $t$ is being measured in seconds. \par
These parameters $z_n$ are especially useful if one wants to solve 
a scale-invariant field equation as we are going to do below. 
As usual, we  define a field equation to be scale-invariant if for any of its solutions 
$g_{ij}$ and any real constant $c$, also the homothetically related metric 
$e^{2c} \, g_{ij}$ represents a solution. \par
Our parameters $z_n$ are related to the more  usual notation, see \cite{m30}, 
\cite{m31}, \cite{m32},  \cite{m33},  and \cite{m34}, as follows: 
\be%2.4
z_2 = \frac{\ddot a \, a}{\dot a^2} = -q \, , \quad z_3 = j \, , \quad z_4 = -k \, . 
\ee
Here, $q$ is the deceleration parameter, $j$ the jerk parameter, and
$k$ the kerk parameter. The notion statefinder parameter refers to the pair $(j, \, s)$, 
where $s$ is defined for $q \ne 1/2$ via 
\be%2.5 
s = \frac{j - 1}{3(q - 1/2)} \, .
\ee
\par
Next, we give some relations between the parameters $z_n$ and the Hubble
parameter $h$: Solving eq. (2.3) for $a^{(n)}$ we get 
\be%2.6
a^{(n)} =   \frac{z_n \,  \dot a^n}{a^{n-1}}\, .
\ee
The temporal derivative of eq. (2.6) has the same l.h.s. as 
eq. (2.6) with $n$ replaced by $n+1$. Equating the related r.h.sides we get
\be%2.7
z_{n+1} = \dot z_n/h + z_n \, (n \, z_2 + 1 - n) \, .
\ee
From eqs. (2.2) and (2.4) we easily deduce $z_2$ in dependence of $h$: 
\be%2.8
z_2 = -q = 1 -  \frac{d}{dt} \left(\frac{1}{h} \right) = 1 + \dot h / h^2 \, , 
\ee
and for $n \ge 3$ we can iteratively deduce $z_n$ in dependence of $h$
with eq. (2.7), the next two terms being
\be%2.9
z_3 = j = \frac{\dot z_2}{h} + z_2(2z_2 - 1) = 1 + 3 \dot h/h^2 + \ddot h/h^3
\ee
and
$$
z_4 = -k = 1 + \frac{6 \dot h}{h^2} + \frac{4 \ddot h}{h^3} +
\frac{3 \dot h^2}{h^4} + \frac{1}{h^4} \frac{d^3h}{dt^3} \, .
$$
So we get the relation to the  other set of dimensionless constants, see 
eq. (3.12) of  \cite{sc136}
\be%2.10  
\varepsilon_p =   \frac{d^p h}{dt^p} \cdot  h^{-p-1} \, .
\ee
This leads to $z_1=   \varepsilon_0 =1$, $z_2= 1 +   \varepsilon_1$, 
 and $z_3 = 1 + 3   \varepsilon_1 +  \varepsilon_2$. \par
If we take the logarithmic cosmic scale factor  $\a$ (see eq. (2.2)) as 
new time coordinate we can rewrite eq. (2.7) as
\be%2.11
z_{n+1} = \frac{d z_n}{d \a} + z_n \, (n \, z_2 + 1 - n) \, .
\ee
After some reformulation we also get as metric 
\be%2.12
ds^2 = \frac{d\a^2}{h(\a)^2} - e^{2\a} \left(dx^2 + dy^2 + dz^2  \right) 
\ee
with the scale invariant parameters being
\be%2.13
q = - 1 - \frac{1}{h} \cdot \frac{dh}{d\a}\, ,
\ee
from eq. (2.8), and from eqs. (2.9.) and (2.10)
\be%2.14
j = 2q^2 + q - \frac{dq}{d \a}
\ee
and 
\be%2.15
k = 3jq + 2j - \frac{dj}{d \a} \, .
\ee
We will apply  these formulas in section 4.

\section{Gauss-Bonnet lagrangian}%3
\setcounter{equation}{0} 
For 2-dimensional space-times, lagrangians of the type $f(R)$  have been discussed e.g.
in \cite{sc57}, \cite{mx23}, \cite{sc78}, and  \cite{mx24}. In  \cite{sc57}, the lagrangian 
\be%3.1
f(R) = R^{k + 1}
\ee
was shown to lead to non-trivial classical results even in the limit
$k \to 0$. In  \cite{sc78}, this limit was shown to produce the 
same field equation as the lagrangian
\be%3.2
f(R) = R \cdot \ln R  \, .
\ee
This property is related to the fact that $\int R \sqrt{g} d^2 x$ is a 
topological invariant related to the genus of the space. \par
Similarly, in \cite{mx24}, the integrand $R$ was kept constant,  but instead,
the dimension of space-time was formally defined as $2 + \epsilon$, and 
the limit $\epsilon  \to 0$  was discussed. \par
We now want to transfer  this idea to the set of  4-dimensional space-times.
 To this end we consider a general function $F(G)$ with $G$ from eq. (1.2)
 as integrand of  the action. The full field equations are given e.g. in eq. (5) of
 \cite{m21}  using the notation  $F_G = dF(G)/dG$,  
\bea%3.3
0 = \frac{1}{2}g^{ij}F(G) - 2 F_G R R^{ij} + 4 F_G R_k^ i R^{kj}
 - 2 F_G R^{iklm} R^j_{\  klm}  \nonumber  \\
 - 4 F_G R^{iklj} R_{kl} + 2RF_G^{;ij}
- 2 g^{ij}R \Box F_G - 4 R^{ik} F^{;j}_{G;k}  \nonumber  \\
 - 4 R^{jk} F^{;i}_{G;k} + 4 R^{ij} \Box F_G+ 
4g^{ij} R^{kl} F_{G;kl} - 4 R^{ikjl}  F_{G;kl}  \, .
\eea
The integral $ I_G = \int G \sqrt{-g} d^4 x $  is a topological invariant related to 
the Euler characteristic. Therefore, the function $F(G)=G$ leads to the field equation 
reading $0=0$  trivially fulfilled by all metrics $g_{ij}$, i.e., every space-time 
represents a stationary point of the   action $I_G$. \par
Next, we consider the action
\be%3.4
I = \int F(G) \sqrt{-g} d^4 x
\ee
and ask for its properties if a scale transformation is applied to the metric.
More exactly: What happens with $I$ eq. (3.4) if we replace $g_{ij}$
 by its homothetically equivalent metric $e^{2 \gamma}g_{ij}$ where
$\gamma$ is an arbitrary constant? If $I$ does not change at all by this transformation,
then we call $I$ scale-invariant. $G$ goes over to  $e^{-4 \gamma} G$  by
this transformation, and so, obviously, only $F(G) = c_2 \cdot G$ with a constant
$c_2$   leads to a scale-invariant action $I$. \par
A less trivial question is the following one: Under which conditions, the action 
$I$ eq. (3.4) is almost scale-invariant, i.e., scale-invariant up to adding a multiple
of $I_G$? In other words: Which functions $F(G)$ have the property
that replacing  $g_{ij}$  by $e^{2 \gamma}g_{ij}$ in eq. (3.4) 
 leads to the action $I + k_\gamma \cdot I_G$ with constant $k_\gamma$?
The answer is: Besides the case already discussed above, 
\be%3.5
F(G) = c_1 \cdot G \cdot \ln G + c_2 \cdot G
\ee
with constants $c_1 \ne 0$ and $c_2$ is the complete set of 
solutions.\footnote{For negative values  of $G$, the term $\ln G$ should be 
replaced by $\ln \vert G \vert$. The singularity at $G \to 0$ is a mild one and 
in the models of our   interest, $\vert G \vert$ is positive anyhow.} 
A word to dimensions: the argument of  the  logarithm should be 
dimensionless, so, instead of $\ln G$ we should have written  $\ln(G/G_0)$.
However, a change of the value $G_0$ can be compensated by a redefinition
of the constant $c_2$. As the term with $c_2$ does not contribute to the
field equation, we may put it to zero classically. Dividing everything by $c_1$
we finally get the only interesting remaining almost scale-invariant case to be
\be%3.6
F(G) = G \cdot \ln G\, .
\ee
If we insert eq. (3.6) into eq. (3.3), the following simpler field equation appears:
\bea%3.7
0 = \frac{1}{2}g^{ij}G \cdot \ln G - 2 ( R R^{ij} -2  R_k^ i R^{kj}
+ R^{iklm} R^j_{\  klm}  \nonumber  \\
 +2 R^{iklj} R_{kl} ) \cdot (1 + \ln G) + 2R  (\ln G)^{;ij}
- 2 g^{ij}R \Box (\ln G) - 4 R^{ik} (\ln G)^{;j}_{;k}  \nonumber  \\
 - 4 R^{jk} (\ln G)^{;i}_{;k} + 4 R^{ij} \Box (\ln G)+ 
4g^{ij} R^{kl} (\ln G)_{;kl} - 4 R^{ikjl}  (\ln G)_{;kl}  \, .
\eea
Due to its importance it seems justified to deduce this case by another way:
 Take a small positive parameter $\epsilon$ and define 
\be%3.8           (\ln G)
F_\epsilon (G) = \frac{1}{\epsilon} \cdot \left( G^{1+\epsilon} - G \right)
\ee
which leads to the same vacuum equation as the lagrangian 
$ G^{1+\epsilon}$. Then the limit $\epsilon \to 0 $ in eq. (3.8) exactly leads
to (3.6). Sketch of the proof:  Put $G=e^x $, then 
$$
G^\epsilon = e^{\epsilon x} \approx 1 + \epsilon x = 1 + \epsilon \ln G\, .
$$

\section{Exact Friedman models}%4
\setcounter{equation}{0} 
We apply the notation of section 2, especially metric (2.1) with Hubble 
parameter (2.2) etc. If we start with $a(t) = t^n$ with positive values $n$ and $t$,
we get $h=n/t$, $\a = n \ln t$, $q=(1-n)/n$, $j=(n-1)(n-2)/n^2$,
$k=- (n-1)(n-2)(n-3)/n^3$, $t= e^{\a/n}$, and $h(\a) = n \cdot e^{-\a/n}$. 
The metric can then also be written as 
\be%4.1
ds^2 = \frac{d \a^2}{n^2}  e^{2\a/n} -  e^{2\a } \left(dx^2 + dy^2 + dz^2 \right)\,   .
\ee
This leads to the de Sitter space-time as $n \to \infty$ where $q= -1$, 
$j=1$,  $k= -1$, and $s=0$. Within Einstein's theory and with pressureless matter
 of density $\rho$, the deceleration $q$ is related to the critical density 
$\rho_c$ necessary to close the universe via $2q = \rho/\rho_c$. \par
Using eqs. (6) and (7) of \cite{m21}, or using eq. (3)  of \cite{mx19} we get
\be%4.2
R = 6 \left( \dot h + 2 h^2 \right )\, , \qquad G = 24 \left( \dot h h^2 +  h^4 \right )
\ee
and the vacuum equation following from the action (3.4)  as
\be%4.3
0 = G \cdot F_G - F(G) - 24 \dot G \cdot  h^3 \cdot F_{GG}\,  ,
\ee
where $F_G = dF/dG$ and $F_{GG} = dF_G/dG$.  In comparison with the 
full field equation (3.3), this is a surprisingly simple equation. 
We test the previously discussed
 property as follows: adding $c_2 \cdot G$ to this $F$, the set of solutions to
eq. (4.3) will not change. For non-vanishing $F_{GG}$, i.e., a non-linear
function $F(G)$, eq. (4.3) is of third order in the metric, as it represents 
the constraint equation to the full fourth-order field equation. \par
Now we insert the example $F(G)= G \ln G$ of eq. (3.6) into eq. (4.3) and get
via $F_G = 1 + \ln G$ and $F_{GG} = 1/G$ and after multiplication with 
$G = - 24 h^4 \cdot q$
\be%4.4
0 =  G^2 - 24 \dot G \cdot  h^3 \, .
\ee
The singular case $G=0$ needs an extra consideration: Looking at eq. (4.2) this leads 
to $\dot h = - h^2$, as the case $h=0$ was already excluded earlier. This behaviour 
can be written in the original form (2.1) by $a(t)= t$, i.e. the deceleration vanishes
 identically: $q=0$. \par
Now we look for the remaining solutions of eq. (4.4), i.e., those with  $G \ne 0$. To this end 
we insert eqs. (2.8) and (4.2) into eq. (4.4). The result is the second order equation for $h$
\be%4.5
\left( \dot h h^2 + h^4 \right)^2 = h^3\cdot \frac{d}{dt} \left( \dot h h^2 + h^4 \right)
\ee
reducing via $\frac{\dot q}{h} = \frac{dq}{d \a}$, see eq. (2.2),  to the following 
first-order equation for the deceleration  parameter $q$
\be%4.6
 \frac{dq}{d \a} = 4q + 3q^2\, . 
\ee

The fact, that eq. (4.6) does not contain the Hubble parameter is a 
consequence of the scale-invariance of the field equation. 
By the way, eqs. (2.14) and (4.6) can be combined to $q^2 + 3q + j=0$
 characterizing this field equation. \par
The other solution with constant value  $q$ is $q= -4/3$. Using eq. (2.8)
we get $3 \dot h = h^2$, i.e. $h=-3/t$ and finally $a(t) = 1/t^3$.
These two solutions with constant $q$, i.e. $a(t)= t$ and  $a(t) = 1/t^3$,
 represent themselves self-similar space-times: Multiplying the metric of space-time 
with a constant factor can be compensated by a time-translation. \par
Let us finally come to the case on non-constant $q$ in eq. (4.6).
 Considering solutions as same, if they are related by a scale-transformation, 
  exactly three solutions remain, characterized by
\be%4.7
q(\a) =  -  \frac{4}{3 + 3 \cdot e^{-4\a}} \, , \qquad -\frac{4}{3} < q < 0 
\ee
and
\be%4.8
q(\a) =  -  \frac{4}{3 - 3 \cdot e^{-4\a}}\, ,
\ee
where $\a$ may take all real values  in eq. (4.7), but eq. (4.8) 
 is not defined for $\a=0$ and represents one solution for $\a >0$,
 i.e. $q < -4/3$ and another one for $\a <0$, i.e. $q>0$. 
Coming back to a relation for the scale factor, we get
\be%4.9
q(a) = - \frac{\ddot a \, a}{\dot a^2} =  -  \frac{4}{3 \pm 3 /a^4}\, .
\ee
One of the  two remaining quadratures can still be done in explicit form via
 eq. (2.13), i.e. $ \frac{d(\ln h)}{d \a} = -1-q$, leading to
\be%4.10
h(a) = \frac{c}{a} \cdot \vert a^4 \pm 1 \vert ^{-1/3}
\ee
with a positive constant $c$. The final step to get the function $a(t)$
is then via the  integral
\be%4.11
\int_{a(0)}^{a(t)} \vert a^4 \pm 1 \vert ^{-1/3} da = c \cdot t \, . 
\ee

\section{Conclusion}%5
\setcounter{equation}{0} 
For the lagrangian $L =  G \ln  G$ where $G$, see eq. (1.2),  is the Gauss-Bonnet
curvature scalar we deduced the field equation and solved it completely up to one 
final quadrature eq. (4.11)  in closed form for 3-flat Friedman models using a 
statefinder parametrization. Further we have shown, that among all lagrangians 
$F(G)$ this $L$ is the only one not having the form $G^r$ with a real constant 
$r$ but possessing a scale-invariant field equation. This turns out to be one of its 
analogies to $f(R)$-theories in 2-dimensional space-time.  \par
 Recently, several other modifications of Einstein gravity have been
discussed, see e.g. \cite{d39} for a non-local one, here we 
propose with the arguments given above, as gravitational lagrangian 
\be%5.1
L_g = \Lambda + R + \a R^2 + \b C_{ijkl} C^{ijkl} + \gamma   G \ln  G
\ee
being worth considered in more details than done  up to now. 

\section{Appendix}%6
\setcounter{equation}{0} 
Here we present some decompositions of the Riemann tensor  from the 
geometric  point of view which are implicitly used in the text  above, and 
which may have some interest in themselves  and have other applications, too. \par
In four dimensions,
 the Riemann tensor $R_{ijkl}$  possesses  $4^4 = 256$ real components. 
By use  of  the known symmetries, this  figure reduces to  $20$, but this 
 twenty-dimensional space is even harder to imagine. Example: To work
with the field equation (3.3) it is necessary to know, that in four dimensions,
$$
C^{iklm} C_{jklm} = \frac{1}{4} \  \delta^i_j  \ C^{gklm} C_{gklm}
$$
and how this can be used for evaluating analogous terms with the Riemann tensor. \par
Below, we will present four  different possibilities  how to arrange  this set of 
components to get a better understandable  system. \par
The Riemann tensor $R_{ijkl}$ of  a space-time of dimension $n \ge 3$
 can be decomposed according to several different criteria: 

\noindent 1. The usual one into the Weyl tensor $C_{ijkl}$  plus a term containing  the 
Ricci tensor $R_{ij}$  plus a term  containing  the Riemann curvature scalar $R$. 

\noindent 2. Two trace-less parts plus the trace.

\noindent 3. The  Weyl tensor plus only one additional term. 

\noindent  4. Two divergence-free parts  plus the trace. 

We use  the following two properties of the Riemann tensor
\be%6.1
R_{ijkl}= - R_{ijlk}\, ,  \qquad R_{ijkl} = R_{klij} \, . 
\ee
The Ricci tensor is the trace of the Riemann tensor: $R_{ij}= g^{kl}R_{ikjl}$,
 where $g_{kl}$ denotes the metric of the space-time, and the Riemann curvature 
scalar is the trace of the Ricci tensor $R = g^{kl}R_{kl}$.
The sign conventions are defined such that in Euclidean signature, the 
curvature scalar of the standard sphere is positive.  \par
For any symmetric tensor $H_{ij}$ we define another tensor  $H^\ast_{ijkl}$ via
\be%6.2
H^\ast_{ijkl} =H_{ik} g_{jl} + H_{jl} g_{ik}-
 H_{il} g_{jk}- H_{jk} g_{il} \, .
\ee
Then the tensor  $H^\ast_{ijkl}$ automatically fulfils the identities
 eq. (6.1). For the special case $H_{ij} = g_{ij}$  we get the simplified form
\be%6.3
g^\ast_{ijkl} = 2 g_{ik} g_{jl} - 2 g_{il} g_{jk} \, .
\ee

\subsection{The usual decomposition}%6.1

The Weyl tensor $C_{ijkl}$ is the trace-less  part of the Riemann tensor, 
i.e. $g^{ik} \, C_{ijkl} = 0$. It vanishes identically for $n=3$.
Using the notation of eqs. (6.2) and (6.3) we make the ansatz 
\be%6.4
R_{ijkl}=C_{ijkl} + \a R^\ast_{ijkl} + \b R \,  g^\ast_{ijkl} \, .
\ee
Then the coefficients $\a$ and $\b$ have to be specified such that 
the trace-lessness condition for the Weyl tensor becomes an identity. This
 condition determines the coefficients $\a$ and $\b$ uniquely, and  the result is: 
\be%6.5
\a = \frac{1}{n-2}\qquad {\rm and}  \qquad   \b = \frac{-1}{2(n-1)(n-2)}  \, .
\ee
Thus, we get the usual formula
\bea
R_{ijkl}=C_{ijkl} + \frac{1}{n-2} \left( R_{ik} g_{jl} + R_{jl} g_{ik}-
 R_{il} g_{jk}- R_{jk} g_{il} \right ) \nonumber  \\
 - \,  \frac{1}{(n-1)(n-2)}   R \left(   g_{ik} g_{jl} -  g_{il} g_{jk} \right)  
\nonumber \, .
\eea

\subsection{The  decomposition using trace-less parts}%6.2
 In distinction to the previous subsection, we now perform  a more 
 consequent decomposition into trace and trace-less parts. To this 
end we define $S_{ij}$ as the trace-less part of the Ricci tensor, 
i.e. $g^{ij} S_{ij} =0$ with $S_{ij}= R_{ij} + \kappa  R g_{ij}$
 possessing the unique solution $\kappa = - 1/n$, i.e., 
$ S_{ij}= R_{ij} -  R \,  g_{ij}/n$. Then  the analogous equation to eq. (6.4) is 
\be%6.6
R_{ijkl}=C_{ijkl} + \g S^\ast_{ijkl} + \d R \,  g^\ast_{ijkl} \, .
\ee
This  becomes a correct  identity if and only if 
\be%6.7
\g = \frac{1}{n-2}   \qquad {\rm and} \qquad  \d = \frac{1}{2n(n-1)} \, .
\ee
So  we get 
\bea
R_{ijkl}=C_{ijkl} + \frac{1}{n-2} \left( S_{ik} g_{jl} + S_{jl} g_{ik}-
 S_{il} g_{jk}- S_{jk} g_{il} \right ) \nonumber  \\
 + \,  \frac{1}{n (n-1)}   R \left(   g_{ik} g_{jl} -  g_{il} g_{jk} \right)  \nonumber  \, .
\eea

\subsection{Decomposition into two parts}%6.3
Let us define a tensor $L_{ij}= R_{ij} + \z R \,  g_{ij}$
such that a parameter $\e$ exists which makes 
\be%6.8
R_{ijkl}=C_{ijkl} + \e L^\ast_{ijkl}
\ee
becoming a true identity. It turns out that this is possible if and only if 
\be%6.9
\z = \frac{-1}{2(n-1)}   \qquad {\rm and} \qquad  \e = \frac{1}{n-2}\, .
\ee
Thus,  we can write $L_{ij}= R_{ij} -    \frac{1}{2(n-1)}  R \,  g_{ij}$
and 
\be%6.10 
R_{ijkl}=C_{ijkl} +  \frac{1}{n-2}  \left( L_{ik} g_{jl} + L_{jl} g_{ik}-
 L_{il} g_{jk}- L_{jk} g_{il} \right) \, .
\ee

\subsection{Decomposition into divergence-free parts}%6.4
Now, besides the  identities from eq. (6.1),  we also use identities  involving 
 the covariant derivatives, denoted by a semicolon,  of the Riemann tensor.  The 
Bianchi identity reads 
$$
R_{ijkl;m} + R_{ijlm;k} + R_{ijmk;l}=0 \, . 
$$
Its trace can be obtained by transvection with $g^{ik}$ and reads
\be%6.11
R_{jl;m} + R^i_{\ jlm;i} - R_{jm;l}=0 \, . 
\ee
It should be mentioned, that the transvection with respect to other pairs of 
 indices does not lead to further identities. The Einstein $E_{ij}$ tensor is defined 
as $E_{ij} = R_{ij} + \lambda \,  R \, g_{ij}$,  where $\lambda$ has to be chosen 
such that  the  Einstein tensor is divergence-free, i.e., $ E^{i}_{\ j ;i} = 0$.
Using the  trace of eq. (6.11)  (again, there is essentially only one 
such trace), namely $2R^i_{\ l;i} - R_{;l} =0$, we uniquely get $\lambda = -1/2$, 
i.e. the Einstein tensor is  $E_{ij} = R_{ij} - \frac{1}{2} R \, g_{ij}$. \par
With the ansatz 
\be%6.12
R_{ijkl} = W_{ijkl} + \h E^\ast_{ijkl} + \t R \,  g^\ast_{ijkl}
\ee
 it holds: The coefficients $\h$ and $\t$ are uniquely determined  by the requirements 
that eq. (6.12) is an identity, and  the divergence of the tensor  $ W_{ijkl}$ vanishes: 
$W^{i}_{\ jkl ;i} =0$. We get uniquely  the following values of  the constants: 
$\h = 1$ and $  \t = \frac{1}{4}$.  Then
\bea
R_{ijkl}=W_{ijkl} +  E_{ik} g_{jl} + E_{jl} g_{ik}- 
 E_{il} g_{jk}- E_{jk} g_{il}  \nonumber  \\
 + \,  \frac{1}{2}   R \left(   g_{ik} g_{jl} -  g_{il} g_{jk} \right)  
\eea
defines a decomposition of the Riemann curvature tensor 
into the divergence-free tensors $W_{ijkl}$, $E_{ij}$, $g_{ij}$
and the scalar $R$. \par
It should be mentioned, that for every  $n>2$, the four tensors
$R_{ij}$, $S_{ij}$, $L_{ij}$, and $E_{ij}$  represent four different tensors. 
And it is a remarkable fact, that the coefficients in $E_{ij}$  and in
eq. (6.13)  do not depend on the dimension $n$.

\end{document}